\documentclass[12pt]{article}
\usepackage{amsfonts}
\usepackage{amsmath}
\usepackage{graphicx}

\newcounter{counter1}

\newcounter{counter4}
\newtheorem{theorem}{Theorem}

\newtheorem{corollary}[counter4]{Corollary}

\newtheorem{lemma}[counter1]{Lemma}

\newenvironment{proof}[1][Proof]{\noindent\textbf{#1.} }{\ \rule{0.5em}{0.5em}}
\input{tcilatex}

\begin{document}

\title{{\Large \textbf{Soliton Solutions on Noncommutative Orbifold $%
T^{2N}/G $}}}
\author{Hui Deng \thanks{%
Email:hdeng\_phy@yahoo.com.cn}, \hspace{5mm} Bo-Yu Hou \thanks{%
Email:byhou@nwu.edu.cn}, \hspace{5mm} Guo-Fang Shi\thanks{%
Email:shiguofang@eyou.com}, \and Kang-Jie Shi\thanks{%
Email:kjshi@nwu.edu.cn}, \hspace{5mm}Rui-Hong Yue\thanks{Email:rhyue@nwu.edu.cn}\hspace{5mm} Hua-Hui Xiong\thanks{%
Email:jimharry@eyou.com} \\
{\footnotesize Institute of Modern Physics, Northwest University,}\\
{\footnotesize Xi'an, 710069, P. R. China}}
\maketitle

\begin{abstract}
In this paper, we construct the common eigenstates of "translation" operators $%
\{U_{s}\}$ and establish the generalized $Kq$ representation \cite{kq1,kq2}%
on integral noncommutative torus $T^{2N}$. We then study the finite rotation
group $G$ in noncommutative space as a mapping in the $Kq$ representation
and prove a Blocking Theorem. We finally obtain the complete set of
projection operators on the integral noncommutative orbifold $T^{2N}/G$ in
terms of the generalized $Kq$ representation. Since projectors are soliton
solutions on noncommutative space in the limit $\alpha ^{\prime
}B_{ij}\rightarrow \infty (\Theta _{ij}/\alpha ^{\prime }\rightarrow 0)$, we
thus obtain all soliton solutions on that orbifold $T^{2N}/G.$
\end{abstract}

\vspace{1cm}
 \medskip \medskip\

\section{Introduction}

The idea about the noncommutative space-time coordinates occurred rather
long ago \cite{noncommutative}. A lot of work develops the idea extensively
from the perspectives of both mathematics and physics\cite{1,2,3}. In the
past few years, it has been shown that some noncommutative gauge theories
can be embedded in string theories\cite{6,connes,witten}. Seiberg and Witten%
\cite{witten} pointed out that in the presence of constant $B$- field, the
effective action of string theory can equivalently be described in terms of
noncommutative gauge theory. Another intriguing finding about noncommutative
field theory is the $UV/IR$ mixing arising from noncommutativity of spacetime%
\cite{bib:uv1,bib:uv2}. Noncommutative geometry can also be applied to
condensed matter physics. The currents and density of a system of electrons
in a strong magnetic field may be described by a noncommutative quantum
field theory \cite{gubser,poly,hell}. The connection between a finite
quantum Hall system and a noncommutative Chern-Simon Matrix model first
proposed by \cite{hall20} was further elaborated in papers \cite%
{hall21,hall22}. Many papers are concentrated on the research for the
related questions about the quantum Hall effect \cite{hall1}-\cite{bib:hall9}%
.

Solitons in various noncommutative theories have played a central role in
understanding the physics of noncommutative theories and certain aspects of
string theories. One of the reasons why the quantum Hall effect has received
so many concerns is that the quantum Hall effect provides us with a
practical objection embodying rich relations among soliton and
noncommutative theory and string theory \cite%
{bib:hall2,hall6,bib:hall9,bib:hall10}. In string theory, the existence and
form of these classical solutions known as solitons are fairly independent
of the details of the theory, making them useful to probe the string
behavior. Soliton solutions and study of integrable systems in context of
noncommutative space have attracted a lot of interests\cite{14}-\cite{9}.
Noncommutative solitons are comprehended as D-branes in string field theory
with a background $B$ field. Many of Sen's conjectures \cite{18,19}
regarding tachyon condensation in string field theory have been beautifully
confirmed using properties of noncommutative solitons.

String theory demands that there are some compactified dimensions in the
target space. Connes, Douglas and Schwartz studied the question of
compactification on noncommutative tori of two kinds of Matrix models in
terms of noncommutative geometry \cite{connes}. To study the soliton
solution on noncommutative compactified space is instructive. Although
Derrick's theorem forbids the existence of soliton solution in 2+1
dimensional commutative scalar field theory\cite{derrick}, however
Gopakumar, Minwalla and Strominger found projectors can be used to construct
the soliton solutions in the noncommutative space\cite{strominger}. Thus to
study projection operators on various noncommutative space is endowed by
direct physical meaning. Reiffel \cite{Rieffel} constructed the complete set
of projection operators on the noncommutative torus. On the basis Boca
studied the projection operators on noncommutative orbifold \cite{Boca}
obtaining important results and showed the well-known example of projection
operator for $T^{2}/Z_{4}$ in terms of the elliptic function. Soliton
solutions in noncommutative gauge theory were introduced by Polychronakos in
\cite{soliton}. Martinec and Moore in their important article deeply studied
soliton solutions namely projectors on a wide variety of orbifolds, and the
relation between physics and mathematics in this area \cite{Martinec}.
Gopakumar, Headrick and Spradlin gave a rather apparent method to construct
the multi-soliton solution on noncommutative integrable torus with generic $%
\tau $\cite{GHS}. We have shown projection operators of a manifest covariant
form on noncommutative orbifold $T^{2}/Z_{4}$\cite{Deng}$.$

The noncommutative spaces in string theory origin from the D-brane on which
the bottom of open string moves. In general D-brane is high dimensional. So
the study of soliton solutions in compactified high dimensional spaces is
very helpful for us to understand the properties of D-brane. In solid
physics the 3-dimensional coordinates space together with the dual momenta
space constitutes a 6-dimensional noncommutative phase space. Since the
spacial objects studied in solid physics possess periodicities, the phase
space must be also a lattice space. There has been some work to do with the
lattice model on high dimensional space. For example, Dai and Song defined
hypercubic group in any dimensions and carried out detailed calculation on
the structure and representation of the four dimensional cubic group $O_{4}$
and its double group \cite{song}$.$In this paper we study the general
solutions of projectors in high dimensional noncommutative space $T^{2N}/G$
in the integral case. Concretely, for the case of $N=3$, what we are
studying is the projector invariant under the rotation which keeps the
six-dimensional phase-space lattice invariant and might be useful in solid
state physics.

This paper is organized as following: in Section 2 we define operators in
the noncommutative space $T^{2N}/G$, where $G$ is a finite subgroup of
rotation group $SO(2N)$. We can prove that in general the problem in high
dimensional noncommutative space doesn't degenerate into the case of direct
product of two dimensions. Thus it is necessary to directly deal with the
questions in the high dimensional case. In Section 3, we establish common
eigenstates $\left\{ \left\vert \vec{B},\vec{q}\right\rangle \right\} $ of
the wrapping operators $\left\{ U_{s}\right\} $, namely the generalized $Kq$
representation in the high dimensional case. In Section 4, we explore the
properties of base vectors such as orthogonality, completeness and
quasi-periodicity. In Section 5, we study the transformation of $\left\vert
\vec{B},\vec{q}\right\rangle $ under action of rotation group $G$ and then
prove the Blocking Theorem and related propositions. Finally, in the
integral case we construct the complete set of projection operators in the
noncommutative orbifold $T^{2N}/G$ in Section 6.

\section{Operators On the Noncommutative $T^{2N}/G$}

Give a set of hermiltian operators $\left\{ x_{i}\right\} ,i=1,2,\cdots ,2N$

\begin{equation}
\left[ x_{i},x_{j}\right] =i\theta _{ij},
\end{equation}%
where $\left\{ \theta _{ij}\right\} $ is an antisymmetric real matrix. The
noncommutative space $R^{2N}$ is formed by the $Tailor$ series of $x_{i}.$
We can construct unitary operators, under action of which, $x_{j}$ undergoes
"translation " transformation%
\begin{equation}
U_{s}=e^{iC_{js}x_{j}}~,~~~~~~~~~~s=1,2,\cdots ,2N,
\end{equation}%
\begin{equation}
x_{j}\longrightarrow x_{j}^{\prime }=U_{s}^{-1}x_{j}U_{s}=x_{j}+d_{js},
\end{equation}%
where $d_{js}=\left[ x_{j},iC_{j^{\prime }s}x_{j^{\prime }}\right]
=-C_{j^{\prime }s}\theta _{jj^{\prime }}.$ In this paper we only consider
the case in which the translations don't degenerate, namely $\det d_{sj}\neq
0.$ We have
\begin{equation}
\det \theta _{jj^{\prime }}\neq 0,~~\det C_{j^{\prime }s}\neq 0.  \label{3}
\end{equation}%
The "translations" in terms of $\left\{ U_{s}\right\} $ generate a set of
lattice in the $R^{2N}$ space$.$ We define the set of all the $Tailor$
series which commute with $\left\{ U_{s}\right\} $ as noncommutative torus $%
T^{2N}$. It is verified that they are composed of the Laurent series of $%
\{u_{s}\},s=1,2,\cdots ,2N$, where
\begin{equation}
u_{s}=e^{ic_{js}x_{j}}~,~~~~~~~~s=1,2,\cdots ,2N,
\end{equation}%
commuting with all the $U_{s}$
\begin{equation}
\left[ U_{s},u_{s^{\prime }}\right] =0,~~~s,s^{\prime }=1,2,\cdots ,2N
\end{equation}%
Now we consider a finite subgroup $G$ of $SO(2N),$ $G=\{R_{k}\}$:%
\begin{equation}
R_{k}:x_{j}\rightarrow x_{j}^{\prime }=\left( \Re _{k}\right)
_{jl}x_{l}=R_{k}^{-1}x_{j}R_{k}
\end{equation}%
\begin{equation}
U_{s}\rightarrow U_{s}^{\prime }=e^{iC_{js}x_{j}^{\prime }}=e^{iC_{js}\left(
\Re _{k}\right) _{jl}x_{l}}.
\end{equation}%
In this paper we study the case of any $s,$ we always have
\begin{equation}
U_{s}^{\prime }\equiv R_{k}^{-1}U_{s}R_{k}=\eta _{sk}\prod_{s^{\prime
}}\left( U_{s^{\prime }}\right) ^{K_{ss^{\prime }}},~~K_{ss^{\prime }}\in
\mathbb{Z}
\left( integer\right)  \label{53}
\end{equation}
\begin{equation*}
\det \left( K_{ss^{\prime }}\right) =1.
\end{equation*}%
Namely under action of $R_{k}$, $U_{s}$ changes according to an integer
matrix $K.$ The group $G$ doesn't change the noncommutative torus. The set
of operators which commute with $\left\{ U_{s}\right\} $ and are invariant
under action of\ $\left\{ R_{k}\right\} $ forms noncommutative orbifold $%
T^{2N}/G.$ When all the $U_{s}$ commute with each other, we call the
noncommutative torus $T^{2N}$ integral. In this paper we study the
projectors $P$ on the integral noncommutative orbifold $T^{2N}/G$, namely
the operators satisfying:%
\begin{equation}
P^{2}=P,
\end{equation}%
\begin{equation}
\left[ U_{s},P\right] =0,
\end{equation}%
\begin{equation}
\left[ R_{k},P\right] =0.
\end{equation}

\section{The Common Eigenvectors of the Wrapping Operators $\left\{
U_{s}\right\} $}

In this section, we construct common eigenstates of the wrapping operators $%
\left\{ U_{s}\right\} $ on the integral $T^{2N}.$ Set
\begin{equation}
C_{js}x_{j}=y_{s},
\end{equation}%
we have
\begin{equation}
U_{s}=e^{iy_{s}}.
\end{equation}%
Due to the mutual commutativity of $\left\{ U_{s}\right\} $, we get
\begin{equation}
\left[ y_{s},y_{s^{\prime }}\right] =2\pi il_{ss^{\prime }},  \label{1}
\end{equation}%
where $l_{ss^{\prime }\text{ }}$is an integer number. It can be verified
that in the case of $\det \left( l_{ss^{\prime }\text{ }}\right) \neq 0,$ $%
l_{ss^{\prime }\text{ }}\in
\mathbb{Z}
,$ which is the consequence of (\ref{3})$,$ we can always expand $y_{s}$ as
the linear combination of the operators $\left\{ \hat{p_{l}},\hat{q}%
_{l}\right\} ,l=1,2,\cdots N$
\begin{equation}
y_{s}=\sum_{l=1}^{N}\alpha _{ls}\hat{p_{l}}+\sum_{l=1}^{N}\beta _{ls}\hat{q}%
_{l}\equiv \vec{\alpha}_{s}\cdot \hat{\vec{p}}+\vec{\beta}_{s}\cdot \hat{%
\vec{q}},  \label{2}
\end{equation}%
such that $\alpha _{sl}$ are integer numbers, $\beta _{sl}$ are rational
numbers times $2\pi $ and $\left\{ \hat{p_{l}},\hat{q}_{l}\right\} $
satisfies the following canonical commutation relations:
\begin{equation}
\left[ \hat{q}_{j},\hat{p}_{j^{\prime }}\right] =i\delta _{jj^{\prime }},~~~%
\left[ \hat{q}_{j},\hat{q}_{j^{\prime }}\right] =\left[ \hat{p}_{j},\hat{p}%
_{j^{\prime }}\right] =0.
\end{equation}%
The N-dimensional vectors $\vec{\alpha}_{s}$ and $\vec{\beta}_{s}$ are
defined via (\ref{2}). Eq. (\ref{1}) and (\ref{2}) imply%
\begin{equation}
\vec{\beta}_{s}\cdot \vec{\alpha}_{s^{\prime }}-\vec{\beta}_{s^{\prime
}}\cdot \vec{\alpha}_{s}=2\pi l_{ss^{\prime }},~~~l_{ss^{\prime }}\in
\mathbb{Z}  \label{37}
\end{equation}%
Since $\alpha _{sj}$ are integers, from (\ref{3}) it can be shown that in
the N-dimensional space for the 2N vectors $\left\{ \vec{\alpha}_{s}\right\}
$ we can always find N vectors $\left\{ \vec{a}_{j}\right\} $ which satisfy
the following conditions,
\begin{equation}
\vec{\alpha}_{s}=\sum_{j=1}^{N}\bar{Z}_{js}\vec{a}_{j};  \label{6}
\end{equation}%
\begin{equation}
\vec{a}_{j}=\sum_{s=1}^{2N}Z_{sj}\vec{\alpha}_{s},  \label{4}
\end{equation}%
where $\bar{Z}_{js},Z_{sj}\in
\mathbb{Z}
\mathbf{~}\left( \text{standing for integer}\right) ,$ $s=1,2,\cdots
,2N,j=1,2,\cdots ,N.$ Note that the solution of $Z_{sj}$ in (\ref{4}) is not
unique, we can arbitrarily choose one from them and define
\begin{equation}
\vec{\gamma}_{j}\equiv \sum_{s=1}^{2N}Z_{sj}\vec{\beta}_{s},  \label{5}
\end{equation}%
We then have from (\ref{37}),(\ref{4}) and (\ref{5})%
\begin{equation*}
\vec{\beta}_{s}\cdot \vec{a}_{j}-\vec{\gamma}_{j}\cdot \vec{\alpha}_{s}=2\pi
\bar{l}_{js}~;
\end{equation*}%
\begin{equation}
\vec{\gamma}_{j}\cdot \vec{a}_{j^{\prime }}-\vec{\gamma}_{j^{\prime }}\cdot
\vec{a}_{j}=2\pi \bar{\bar{l}}_{jj^{\prime }}.  \label{10}
\end{equation}%
with $\bar{l}_{js},\bar{\bar{l}}_{jj^{\prime }}\in
\mathbb{Z}
.$

Now we start constructing the common eigenstates of $\left\{ U_{s}\right\} $%
. Consider the lattice vectors generated by $\left\{ \vec{a}_{j}\right\} $
in the N-dimensional space:
\begin{equation}
\vec{m}=\vec{r}_{\left\{ m_{j}\right\} }=\sum_{j=1}^{N}m_{j}\vec{a}%
_{j},~~~m_{j}\in
\mathbb{Z}
.  \label{8}
\end{equation}%
From (\ref{6}) any vector generated by linear combination of $\vec{\alpha}%
_{s}$ with integral coefficients is located on the lattice. Now we set the
common eigenstates of $\left\{ U_{s}\right\} $ as follows,
\begin{equation}
\left\vert \vec{B},\vec{q}\right\rangle =\sum_{\vec{m}}e^{i\left( \vec{m}%
^{T}A\vec{m}+\vec{m}\cdot \vec{B}\right) }\left\vert \vec{q}+\vec{m}%
\right\rangle ,
\end{equation}%
where $\vec{B},\vec{q},\vec{m}$ are all N-dimensional vectors, and $A$ is an
$N\times N$ real matrix, the sum goes over all the lattice sites. $%
\left\vert \vec{q}+\vec{m}\right\rangle $ is common eigenstate of the
coordinate operators $\left\{ \hat{q}_{j}\right\} $,
\begin{equation}
\hat{q}_{j}\left\vert \vec{q}+\vec{m}\right\rangle =\left( \vec{q}+\vec{m}%
\right) _{j}\left\vert \vec{q}+\vec{m}\right\rangle .
\end{equation}%
Note:%
\begin{equation*}
\left( \vec{m}\right) _{j}=\sum_{l=1}^{N}m_{l}\left( \vec{a}_{l}\right)
_{j}\neq m_{j}.
\end{equation*}

We require%
\begin{equation}
U_{s}\left\vert \vec{B},\vec{q}\right\rangle =\lambda _{s}\left( \vec{B},%
\vec{q}\right) \left\vert \vec{B},\vec{q}\right\rangle  \label{39}
\end{equation}%
obtaining%
\begin{eqnarray}
&&e^{-\frac{i}{2}\left( \vec{\alpha}_{s}\cdot \vec{\beta}_{s}\right) }e^{i%
\vec{\alpha}_{s}\cdot \hat{\vec{p}}}e^{i\vec{\beta}_{s}\hat{\cdot \vec{q}}%
}\left\vert \vec{B},\vec{q}\right\rangle \\
&=&e^{-\frac{i}{2}\left( \vec{\alpha}_{s}\cdot \vec{\beta}_{s}\right) }\sum_{%
\vec{m}}e^{i\left( \vec{m}^{T}A\vec{m}+\vec{m}\cdot \vec{B}\right) }e^{i\vec{%
\beta}_{s}\cdot \left( \vec{q}+\vec{m}\right) }\left\vert \vec{q}+\vec{m}-%
\vec{\alpha}_{s}\right\rangle  \label{38}
\end{eqnarray}%
Since $\vec{m}-\vec{\alpha}_{s}$ is also at the lattice site due to (\ref{6}%
), we may rewrite the right-hand side of (\ref{38}) as
\begin{eqnarray*}
U_{s}\left\vert \vec{B},\vec{q}\right\rangle &=&e^{-\frac{i}{2}\left( \vec{%
\alpha}_{s}\cdot \vec{\beta}_{s}\right) }\sum_{\vec{m}^{\prime }}e^{i\left(
\left( \vec{m}^{\prime }+\vec{\alpha}_{s}\right) ^{T}A\left( \vec{m}^{\prime
}+\vec{\alpha}_{s}\right) +\left( \vec{m}^{\prime }+\vec{\alpha}_{s}\right)
\cdot \vec{B}\right) }e^{i\vec{\beta}_{s}\cdot \left( \vec{q}+\vec{m}%
^{\prime }+\vec{\alpha}_{s}\right) }\left\vert \vec{q}+\vec{m}^{\prime
}\right\rangle \\
&=&\lambda _{s}\sum_{\vec{m}}e^{i\left( \vec{m}^{T}A\vec{m}+\vec{m}\cdot
\vec{B}\right) }\left\vert \vec{q}+\vec{m}\right\rangle .
\end{eqnarray*}%
Naturally we demand
\begin{equation}
e^{i\left( \vec{\alpha}_{s}^{T}\left( A+A^{T}\right) \vec{m}+\vec{\beta}%
_{s}\cdot \vec{m}\right) }=\lambda _{s}e^{i\left( \vec{\alpha}_{s}^{T}A\vec{%
\alpha}_{s}-\vec{\alpha}_{s}\cdot \vec{B}-\vec{\beta}_{s}\cdot \vec{q}%
\right) +\frac{i}{2}\vec{\alpha}_{s}\cdot \vec{\beta}_{s}}
\end{equation}%
to be valid for all $\vec{m}$ belonging to the lattice vectors in (\ref{8}).
The rhs of the above equation is independent of the lattice sites, requiring
the condition
\begin{equation}
\vec{\alpha}_{s}^{T}\left( A+A^{T}\right) \vec{m}+\vec{\beta}_{s}\cdot \vec{m%
}\in 2\pi
\mathbb{Z}%
\end{equation}%
should be satisfied for all $\vec{m}.$ Since $\vec{m}$ may be generated by
the integral coefficient linear combination of $\vec{\alpha}_{s},$ this means%
\begin{equation}
\vec{\alpha}_{s}^{T}\left( A+A^{T}\right) \vec{\alpha}_{s^{\prime }}+\vec{%
\beta}_{s}\cdot \vec{\alpha}_{s^{\prime }}\in 2\pi
\mathbb{Z}
\label{29}
\end{equation}%
for all $s,s^{\prime }.$ If the $N\times N$ matrix $A$ satisfies the above
condition, then (\ref{39}) holds with
\begin{equation*}
\lambda _{s}=\lambda \left( s,\vec{B},\vec{q}\right) =e^{i\left( -\vec{\alpha%
}_{s}^{T}A\vec{\alpha}_{s}+\vec{\alpha}_{s}\cdot \vec{B}+\vec{\beta}%
_{s}\cdot \vec{q}-\frac{1}{2}\vec{\alpha}_{s}\cdot \vec{\beta}_{s}\right) }.
\end{equation*}%
Define $g_{s}$ by
\begin{equation}
-\vec{\alpha}_{s}^{T}A\vec{\alpha}_{s}-\frac{1}{2}\vec{\alpha}_{s}\cdot \vec{%
\beta}_{s}=\pi g_{s},~~~g_{s}\in
\mathbb{Z}%
\end{equation}%
\begin{equation}
\lambda _{s}=\left( -1\right) ^{g_{s}}e^{i\left( \vec{\alpha}_{s}\cdot \vec{B%
}+\vec{\beta}_{s}\cdot \vec{q}\right) }.  \label{17}
\end{equation}%
When $A=A^{T\text{ }},$(see (\ref{7}) below) we have $g_{s}\in
\mathbb{Z}
$ from (\ref{29}). In order to solve matrix $A$ satisfying Eq.(\ref{29}), we
consider the following equation:%
\begin{equation}
\vec{a}_{j}^{T}\left( A+A^{T}\right) \vec{a}_{j^{\prime }}+\vec{\gamma}%
_{j}\cdot \vec{a}_{j^{\prime }}=2\pi T_{jj^{\prime }}~~~T_{jj^{\prime }}\in
\mathbb{Z}
\label{30}
\end{equation}%
For the convenience, we define $N\times N$ matrices $H$ and $a$ via
\begin{equation}
H_{jj^{\prime }}=\vec{a}_{j}^{T}\left( A+A^{T}\right) \vec{a}_{j^{\prime }},
\end{equation}%
\begin{equation}
a_{kj}\equiv \left( \vec{a}_{j}\right) _{k},
\end{equation}%
\begin{equation}
H=a^{T}\left( A+A^{T}\right) a.
\end{equation}%
The crucial point is that $H$ must be symmetric. Since we have $\det a\neq 0$
and there exists the inverse of $a$, so long as we could find symmetric
matrix $H$ satisfying the following equation:%
\begin{equation}
H_{jj^{\prime }}+\vec{\gamma}_{j}\cdot \vec{a}_{j^{\prime }}=2\pi
T_{jj^{\prime }}.  \label{9}
\end{equation}%
then Eq.(\ref{30}) can be solved. Actually, we may set
\begin{equation}
2A=\left( a^{-1}\right) ^{T}Ha^{-1}\equiv X.  \label{7}
\end{equation}%
Due to
\begin{equation}
X^{T}=\left( a^{-1}\right) ^{T}H^{T}a^{-1}=X,
\end{equation}%
Eq.(\ref{30}) can be verified. Next we solve matrix $H$:

\begin{enumerate}
\item When $j\leq j^{\prime },$ for an arbitrary integer matrix elements $%
T_{jj^{\prime }}.Let$
\begin{equation*}
H_{jj^{\prime }}=2\pi T_{jj^{\prime }}-\vec{\gamma}_{j}\cdot \vec{a}%
_{j^{\prime }}
\end{equation*}%
Eq.(\ref{9}) is certainly satisfied.

\item When $j>j^{\prime },$ we take%
\begin{equation*}
H_{jj^{\prime }}=H_{j^{\prime }j}=2\pi T_{j^{\prime }j}-\vec{\gamma}%
_{j^{\prime }}\cdot \vec{a}_{j}.
\end{equation*}%
Due to (\ref{10})%
\begin{eqnarray*}
H_{jj^{\prime }}+\vec{\gamma}_{j}\cdot \vec{a}_{j^{\prime }} &=&2\pi
T_{j^{\prime }j}-\vec{\gamma}_{j^{\prime }}\cdot \vec{a}_{j}+\vec{\gamma}%
_{j}\cdot \vec{a}_{j^{\prime }} \\
&=&2\pi T_{j^{\prime }j}+2\pi \bar{\bar{l}}_{jj^{\prime }}=2\pi \left(
T_{j^{\prime }j}+\bar{\bar{l}}_{jj^{\prime }}\right)
\end{eqnarray*}%
Therefore, when $j>j^{\prime }$ we set
\begin{equation*}
T_{jj^{\prime }}=T_{j^{\prime }j}+\bar{\bar{l}}_{jj^{\prime }},
\end{equation*}%
which is an integer.
\end{enumerate}

Eq.(\ref{9}) then holds for all cases. That is%
\begin{equation*}
\vec{a}_{j}^{T}\left( A+A^{T}\right) \vec{a}_{j^{\prime }}+\vec{\gamma}%
_{j}\cdot \vec{a}_{j^{\prime }}\in 2\pi
\mathbb{Z}
.
\end{equation*}%
From (\ref{6}), we obtain
\begin{equation*}
\vec{a}_{j}^{T}\left( A+A^{T}\right) \vec{\alpha}_{s}+\vec{\gamma}_{j}\cdot
\vec{\alpha}_{s}\in 2\pi
\mathbb{Z}
,
\end{equation*}%
By (\ref{10}) we have
\begin{equation*}
\vec{\alpha}_{s}^{T}\left( A+A^{T}\right) \vec{a}_{j}+\vec{\beta}_{s}\cdot
\vec{a}_{j}\in 2\pi
\mathbb{Z}
,
\end{equation*}%
We further have%
\begin{equation*}
\vec{\alpha}_{s^{\prime }}^{T}\left( A+A^{T}\right) \vec{\alpha}_{s}+\vec{%
\alpha}_{s^{\prime }}\cdot \vec{\beta}_{s}\in 2\pi
\mathbb{Z}
.
\end{equation*}

\section{The Properties of Eigenvector $\left\vert \vec{B},\vec{q}%
\right\rangle $}

\subsection{\protect\bigskip Periodicity}

In an N-dimensional space we define N vectors $\left\{ \vec{b}_{j}\right\} $
dual to $\left\{ \vec{a}_{j}\right\} ,$ satisfying

\begin{equation}
\vec{a}_{j}\cdot \vec{b}_{j^{\prime }}=2\pi \delta _{jj^{\prime }}.
\label{43}
\end{equation}%
Due to%
\begin{equation}
\left\vert \vec{B},\vec{q}\right\rangle =\sum_{\vec{m}}e^{i\left( \vec{m}%
^{T}A\vec{m}+\vec{m}\cdot \vec{B}\right) }\left\vert \vec{q}+\vec{m}%
\right\rangle ,
\end{equation}%
and From (\ref{8}) we have%
\begin{equation}
\left\vert \vec{B}+\vec{b}_{j},\vec{q}\right\rangle =\left\vert \vec{B},\vec{%
q}\right\rangle .
\end{equation}%
We also have%
\begin{equation}
\left\vert \vec{B}-\vec{\gamma}_{j},\vec{q}+\vec{a}_{j}\right\rangle =\sum_{%
\vec{m}}e^{i\left( \vec{m}^{T}A\vec{m}+\vec{m}\cdot \left( \vec{B}-\vec{%
\gamma}_{j}\right) \right) }\left\vert \vec{q}+\vec{a}_{j}+\vec{m}%
\right\rangle .  \label{41}
\end{equation}%
On the other hand, since $\vec{a}_{j}$ is on the lattice formed by $\vec{m}$%
, we have
\begin{eqnarray}
\left\vert \vec{B},\vec{q}\right\rangle &=&\sum_{\vec{m}+\vec{a}%
_{j}}e^{i\left( \vec{m}+\vec{a}_{j}\right) ^{T}A\left( \vec{m}+\vec{a}%
_{j}\right) +i\left( \vec{a_{j}}+\vec{m}\right) \cdot \vec{B}}\left\vert
\vec{q}+\vec{m}+\vec{a}_{j}\right\rangle  \notag \\
&=&\sum_{\vec{m}}e^{i\vec{m}^{T}A\vec{m}+i\vec{m}\cdot \vec{B}+i\vec{a}%
_{j}^{T}\left( A+A^{T}\right) \vec{m}+i\vec{a}_{j}^{T}A\vec{a_{j}}+i\vec{a}%
_{j}\cdot \vec{B}}\left\vert \vec{q}+\vec{m}+\vec{a}_{j}\right\rangle .
\label{40}
\end{eqnarray}%
From (\ref{30}), we have%
\begin{equation*}
\vec{a}_{j}^{T}\left( A+A^{T}\right) \vec{m}+\vec{\gamma}_{j}\vec{m}\in 2\pi
\mathbb{Z}.
\end{equation*}%
Comparing (\ref{41}) and (\ref{40}), we have
\begin{equation}
\left\vert \vec{B},\vec{q}\right\rangle =e^{i\left( \vec{a_{j}}^{T}A\vec{%
a_{j}}+\vec{a_{j}}\cdot \vec{B}\right) }\left\vert \vec{B}-\vec{\gamma}_{j},%
\vec{q}+\vec{a}_{j}\right\rangle .
\end{equation}%
So the eigenstate $\left\vert \vec{B},\vec{q}\right\rangle $ has two sets of
quasi-periodicity relations. In the view of 2N-dimensional $\left\{ \vec{B},%
\vec{q}\right\} $ space, the two sets of quasi-periodic vectors are
respectively
\begin{equation}
\left\{ \vec{b}_{j},0\right\} \text{and}\left\{ -\vec{\gamma}_{j},\vec{a}%
_{j}\right\} .  \label{13}
\end{equation}%
\begin{equation}
j=1,2,\cdots ,N  \label{14}
\end{equation}%
Similarly, we can prove that for $s=1,2,\cdots ,2N$%
\begin{equation}
\left\vert \vec{B},\vec{q}\right\rangle =e^{i\left( \vec{\alpha}_{s}^{T}A%
\vec{\alpha _{s}}+\vec{\alpha _{s}}\cdot \vec{B}\right) }\left\vert \vec{B}-%
\vec{\beta}_{s},\vec{q}+\vec{\alpha}_{s}\right\rangle .
\end{equation}

\subsection{Orthogonal Relation}

Consider the unit cell $V_{a}$ composed of $\left\{ \vec{a}_{j}\right\} $
and the unit cell $V_{b}$ composed of $\left\{ \vec{b}_{j}\right\} .$
\begin{equation}
<\vec{B}^{\prime },\vec{q}^{\prime }|\vec{B},\vec{q}>=\sum_{\vec{m}\vec{m}%
^{\prime }}e^{-i\vec{m}^{\prime T}A\vec{m}^{\prime }-i\vec{m}^{\prime }\cdot
\vec{B}^{\prime }+i\vec{m}^{T}A\vec{m}+i\vec{m}\cdot \vec{B}}\delta
^{N}\left( \vec{q}-\vec{q}^{\prime }+\vec{m}-\vec{m}^{\prime }\right) .
\end{equation}%
When $\vec{q}$ and $\vec{q}^{\prime }$ are both inside $V_{a},$ only when $%
\vec{m}=\vec{m}^{\prime },\delta ^{N}\left( \vec{q}-\vec{q}^{\prime }+\vec{m}%
-\vec{m}^{\prime }\right) \neq 0,$ then
\begin{eqnarray}
<\vec{B}^{\prime },\vec{q}^{\prime }|\vec{B},\vec{q}> &=&\sum_{\vec{m}}e^{i%
\vec{m}\cdot \left( \vec{B}-\vec{B}^{\prime }\right) }\delta ^{N}\left( \vec{%
q}-\vec{q}^{\prime }\right)  \notag  \label{11} \\
&=&\delta ^{N}\left( \vec{q}-\vec{q}^{\prime }\right) \sum_{\left\{ m_{j}\in
\mathbb{Z}\right\} }e^{i\sum_{j}m_{j}\vec{a}_{j}\cdot \left( \vec{B}-\vec{B}%
^{\prime }\right) }.
\end{eqnarray}%
Considering the case that $\vec{B}$ and $\vec{B}^{\prime }$ are both inside $%
V_{b},$ we have
\begin{equation*}
\vec{B}-\vec{B}^{\prime }=\sum_{j}\vec{b}_{j}\left( c_{j}-c_{j}^{\prime
}\right) ,~~~-1<c_{j}-c_{j}^{\prime }<1,~~\left\vert \vec{a}_{j}\cdot \left(
\vec{B}-\vec{B}^{\prime }\right) \right\vert <2\pi .
\end{equation*}%
The right-hand side of Eq.(\ref{11}) becomes%
\begin{eqnarray}
&&\delta ^{N}\left( \vec{q}-\vec{q}^{\prime }\right) \prod_{j}\left( 2\pi
\right) \delta \left( \vec{a}_{j}\cdot \left( \vec{B}-\vec{B}^{\prime
}\right) \right)  \notag  \label{16} \\
&=&\delta ^{N}\left( \vec{q}-\vec{q}^{\prime }\right) \prod_{j}\delta \left(
c_{j}-c_{j}^{\prime }\right) .
\end{eqnarray}%
Therefore in the unit cell $V_{a}\times V_{b},$ the different eigenstates $%
\left\vert \vec{B},\vec{q}\right\rangle $ are orthogonal.

\subsection{Completeness}

Consider the integration over unit cell $V_{2}^{\prime }=V_{a}\times V_{b}$
\begin{eqnarray}
\pounds &=&\int_{V_{2}^{\prime }}d^{N}\vec{B}d^{N}\vec{q}\left\vert \vec{B},%
\vec{q}\right\rangle \left\langle \vec{B},\vec{q}\right\vert  \notag \\
&=&\int_{V_{2}^{\prime }}d^{N}\vec{B}d^{N}\vec{q}\sum_{\vec{m}\vec{m}%
^{\prime }}e^{-i\vec{m}^{\prime T}A\vec{m}^{\prime }-i\vec{m}^{\prime }\cdot
\vec{B}+i\vec{m}^{T}A\vec{m}+i\vec{m}\cdot \vec{B}}\left\vert \vec{q}+\vec{m}%
\right\rangle \left\langle \vec{q}+\vec{m}^{\prime }\right\vert .  \label{12}
\end{eqnarray}%
Let $\vec{B}=\sum_{j}\vec{b}_{j}c_{j},$ we have
\begin{equation*}
e^{i\left( \vec{m}-\vec{m}^{\prime }\right) \cdot \vec{B}}=e^{2\pi
i\sum_{j}\left( m_{j}-m_{j}^{\prime }\right) \cdot c_{j}}.
\end{equation*}%
The integration with respect to $\vec{B}$ inside the unit cell $V_{b}$ can
be realized through changing it into the integration with respect to $%
c_{j}\in \left[ 0,1\right) ,$
\begin{equation*}
\int_{V_{b}}d^{N}Be^{i\left( \vec{m}-\vec{m}^{\prime }\right) \cdot \vec{B}%
}=\{_{\upsilon _{b}~~\vec{m}=\vec{m}^{\prime }}^{0~~~\vec{m}\neq \vec{m}%
^{\prime }},
\end{equation*}%
where $\upsilon _{b}$ is the volume of $V_{b}$ in Eq.(\ref{12}). Then we have%
\begin{eqnarray}
\pounds &=&\int_{V_{a}}\upsilon _{b}d^{N}\vec{q}\sum_{\vec{m}}\left\vert
\vec{q}+\vec{m}\right\rangle \left\langle \vec{q}+\vec{m}\right\vert \\
&=&\upsilon _{b}\int_{-\infty }^{\infty }\prod_{j}dq_{j}\left\vert \vec{q}%
\right\rangle \left\langle \vec{q}\right\vert =\upsilon _{b}id.  \notag
\end{eqnarray}%
We then get the completeness condition:%
\begin{equation}
id=\frac{1}{\upsilon _{b}}\int_{V_{2}^{\prime }}d^{N}\vec{B}d^{N}\vec{q}%
\left\vert \vec{B},\vec{q}\right\rangle \left\langle \vec{B},\vec{q}%
\right\vert .  \label{15}
\end{equation}%
Since we have the other set of periodic vectors (\ref{13}) we can also
rewrite the completeness condition and orthogonal condition (\ref{16}) in $%
V_{2}=V_{a}^{\prime }\times V_{b},$ where $V_{a}^{\prime }$ is the unit cell
composed of quasi-periodic vectors (\ref{14}).

\subsection{Degenerate Lattice}

Now we study the degeneracy of eigenvalue $\lambda _{s}\left( \vec{B},\vec{q}%
\right) $ of $U_{s}$ in the space $\left( \vec{B},\vec{q}\right) .$ Consider
a 2N-dimensional vector $\left(
\begin{array}{c}
\vec{\alpha}_{s} \\
\vec{\beta}_{s}%
\end{array}%
\right) \equiv \vec{w}_{s}.$ From (\ref{3}) $\det \left( C_{js}\right) \neq
0 $, we know that $\left\{ \vec{y}_{s}\right\} $ are linearly independent of
each other. We can always find the dual 2N-dimensional vectors $\vec{\tau}%
_{s}$ satisfying
\begin{equation}
\vec{w}_{s}\cdot \vec{\tau}_{s^{\prime }}=2\pi \delta _{ss^{\prime
}},~~s,s^{\prime }=1,2,\cdots ,2N  \label{33}
\end{equation}%
then when
\begin{equation*}
\left(
\begin{array}{c}
\triangle \vec{B} \\
\triangle \vec{q}%
\end{array}%
\right) =\sum_{s}\vec{\tau}_{s}n_{s},~~~n_{s}\in Z,
\end{equation*}%
we have%
\begin{equation}
\vec{\alpha}_{s}\cdot \triangle \vec{B}+\vec{\beta}_{s}\cdot \triangle \vec{q%
}=\vec{w}_{s}\cdot \sum_{s^{\prime }}\vec{\tau}_{s^{\prime }}n_{s^{\prime
}}=\sum_{s^{\prime }}2\pi \delta _{ss^{\prime }}n_{s^{\prime }}=2\pi n_{s}.
\end{equation}%
Then from (\ref{17}), $\left\vert \vec{B},\vec{q}\right\rangle $ and $%
\left\vert \vec{B}+\triangle \vec{B},\vec{q}+\triangle \vec{q}\right\rangle $
are degenerate for the eigenvalues of all $\left\{ U_{s}\right\} .$Consider
the unit cell $\sigma _{00}$ made up of $\left\{ \vec{\tau}_{s}\right\} $ in
the $\left( \vec{B},\vec{q}\right) $ space. The eigenstates of $\left\vert
\vec{B},\vec{q}\right\rangle $ and $\left\vert \vec{B}^{\prime },\vec{q}%
^{\prime }\right\rangle $ which respectively correspond to two different
points $\left(
\begin{array}{c}
\vec{B} \\
\vec{q}%
\end{array}%
\right) $ and $\left(
\begin{array}{c}
\vec{B}^{\prime } \\
\vec{q}^{\prime }%
\end{array}%
\right) $ inside $\sigma _{00}$ are not degenerate. Let us check this in the
following
\begin{equation}
\left(
\begin{array}{c}
\vec{B}^{\prime }-\vec{B} \\
\vec{q}^{\prime }-\vec{q}%
\end{array}%
\right) =\left(
\begin{array}{c}
\triangle \vec{B} \\
\triangle \vec{q}%
\end{array}%
\right) =\sum_{s}\vec{\tau}_{s}\nu _{s},~~~-1<\nu _{s}<1,  \label{19}
\end{equation}%
\begin{equation}
\frac{\lambda _{s}\left( \vec{B}^{\prime },\vec{q}^{\prime }\right) }{%
\lambda _{s}\left( \vec{B},\vec{q}\right) }=e^{i\left( \vec{\alpha}_{s}\cdot
\triangle \vec{B}+\vec{\beta}_{s}\cdot \triangle \vec{q}\right) }=e^{i\vec{w}%
_{s}\cdot \sum_{s^{\prime }}\vec{\tau}_{s^{\prime }}\nu _{s^{\prime
}}}=e^{2\pi i\nu _{s}}.  \label{42}
\end{equation}%
in the region of (\ref{19}) when and only when $\nu _{s}=0,$the right-hand
side of \ref{42} equals to $1.$

We call the lattice generated by $\vec{\tau}_{s}$ in the $\left( \vec{B},%
\vec{q}\right) $ space as degenerate lattice and the unit cell is the
nondegenerate region of $\left\vert \vec{B},\vec{q}\right\rangle $ with
respect to $\left\{ U_{s}\right\} .$

\subsection{The Degree of Degeneracy in the Space of $\left( \vec{B},\vec{q}%
\right) .$}

In the 2N-dimensional $\left( \vec{B},\vec{q}\right) $ space we have three
lattices:

\begin{enumerate}
\item The lattice $L_{1}$ generated by
\begin{equation*}
\left(
\begin{array}{c}
\vec{\beta}_{s} \\
-\vec{\alpha}_{s}%
\end{array}%
\right) =J\left(
\begin{array}{c}
\vec{\alpha}_{s} \\
\vec{\beta}_{s}%
\end{array}%
\right) =J\vec{w}_{s},
\end{equation*}%
here
\begin{equation*}
J=\left(
\begin{array}{cc}
0 & I \\
-I & 0%
\end{array}%
\right) .
\end{equation*}

\item The lattice $L_{2}$ generated by $\left(
\begin{array}{c}
\vec{b}_{j} \\
0%
\end{array}%
\right) $ and $\left(
\begin{array}{c}
-\vec{\gamma}_{j} \\
\vec{a}_{j}%
\end{array}%
\right) .$

\item The lattice $L_{3}$ generated by $\vec{\tau}_{s}.$
\end{enumerate}

Obviously $L_{3}$ has to contain $L_{2}$ because two associated states
connected by the periodic vectors at most differ by a phase factor,
therefore the eigenvalues with respect to $U_{s}$ must be the same, and the
periodic vectors have to be vectors belonging to $L_{3},$we denote the
vectors in $L_{3}$ by $\vec{t}=\sum_{s=1}^{2N}\vec{\tau}_{s}n_{s},~n_{s}\in
\mathbb{Z}
$. On the other hand, we can prove that $J\vec{w}_{s}$ belongs to the
periodic lattice vectors as follows:
\begin{eqnarray}
e^{i\left( \vec{\alpha}\cdot \hat{\vec{p}}+\vec{\beta}\hat{\cdot \vec{q}}%
\right) }\left\vert \vec{B},\vec{q}\right\rangle &=&e^{-\frac{i}{2}\left(
\vec{\alpha}\cdot \vec{\beta}\right) }e^{i\vec{\alpha}\cdot \hat{\vec{p}}%
}e^{i\vec{\beta}\hat{\cdot \vec{q}}}\sum_{\vec{m}}e^{i\left( \vec{m}^{T}A%
\vec{m}+\vec{m}\cdot \vec{B}\right) }\left\vert \vec{q}+\vec{m}\right\rangle
\notag \\
&=&e^{-\frac{i}{2}\left( \vec{\alpha}\cdot \vec{\beta}\right) }e^{i\vec{%
\alpha}\cdot \hat{\vec{p}}}\sum_{\vec{m}}e^{i\vec{m}^{T}A\vec{m}+i\vec{m}%
\cdot \left( \vec{B}+\vec{\beta}\right) }\left\vert \vec{q}+\vec{m}%
\right\rangle  \notag \\
&=&e^{-\frac{i}{2}\left( \vec{\alpha}\cdot \vec{\beta}\right) }e^{i\vec{\beta%
}\cdot \vec{q}}\sum_{\vec{m}}e^{i\vec{m}^{T}A\vec{m}+i\vec{m}\cdot \left(
\vec{B}+\vec{\beta}\right) }\left\vert \vec{q}+\vec{m}-\vec{\alpha}%
\right\rangle  \notag \\
&=&e^{-\frac{i}{2}\left( \vec{\alpha}\cdot \vec{\beta}\right) }e^{i\vec{\beta%
}\cdot \vec{q}}\left\vert \vec{B}+\vec{\beta},\vec{q}-\vec{\alpha}%
\right\rangle  \label{20}
\end{eqnarray}%
However, when taking $\vec{\alpha}=\vec{\alpha}_{s},\vec{\beta}=\vec{\beta}%
_{s},$ on the left hand side of (\ref{20}) equals to $U_{s}\left\vert \vec{B}%
,\vec{q}\right\rangle =\lambda _{s}\left\vert \vec{B},\vec{q}\right\rangle ,$
so $\left(
\begin{array}{c}
\vec{\beta}_{s} \\
-\vec{\alpha}_{s}%
\end{array}%
\right) $ belongs to periodic lattice too, $L_{1}$ is naturally contained in
$L_{2}.$

\begin{equation}
L_{1}\subset L_{2}\subset L_{3}.
\end{equation}

In order to calculate how many points of $L_{2}$ there exist in a unit cell
of $L_{1}$ and how many points of $L_{3}$ in a unit cell of $L_{2}$, we
compare the volumes of three kinds of unit cells and respectively set them
to be $\upsilon _{1},\upsilon _{2}$ and $\upsilon _{3}$ corresponding to the
three kind of unit cells. Let $\Gamma $ and $W$ be the matrices defined as $%
\Gamma _{s^{\prime }s}=\left( \vec{\tau}_{s}\right) _{s^{\prime
}},W_{s^{\prime }s}=\left( \vec{w}_{s}\right) _{s^{\prime }}.$ From (\ref{33}%
) we have
\begin{equation*}
\Gamma _{s^{\prime }s_{1}}W_{s^{\prime }s_{2}}=\vec{\tau}_{s_{1}}\cdot \vec{w%
}_{s_{2}}=2\pi \delta _{s_{1}s_{2}}
\end{equation*}%
to give out%
\begin{equation*}
\Gamma ^{T}W=2\pi I_{2N\times 2N}
\end{equation*}%
and
\begin{equation}
\det \Gamma \cdot \det W=\left( 2\pi \right) ^{2N}\det I=\left( 2\pi \right)
^{2N}
\end{equation}%
Consequently, we have%
\begin{equation}
\upsilon _{3}=\left\vert \det \Gamma \right\vert =\left( 2\pi \right)
^{2N}\left\vert \det W\right\vert ^{-1}
\end{equation}%
In order to calculate the determinant of $W$, we calculate $W^{T}JW$ by (\ref%
{37})
\begin{eqnarray}
&&W^{T}JW  \notag \\
&=&\left(
\begin{array}{cc}
\vec{\alpha}_{1}^{T} & \vec{\beta}_{1}^{T} \\
\vdots & \vdots \\
\vec{\alpha}_{2N}^{T} & \vec{\beta}_{2N}^{T}%
\end{array}%
\right) \left(
\begin{array}{cc}
0 & I \\
-I & 0%
\end{array}%
\right) \left(
\begin{array}{ccc}
\vec{\alpha}_{1} & \cdots & \vec{\alpha}_{2N} \\
\vec{\beta}_{1} & \cdots & \vec{\beta}_{2N}%
\end{array}%
\right)  \notag \\
&=&\left(
\begin{array}{cccc}
-\vec{\beta}_{1}\cdot \vec{\alpha}_{1}+\vec{\alpha}_{1}\cdot \vec{\beta}_{1}
& -\vec{\beta}_{1}\cdot \vec{\alpha}_{2}+\vec{\alpha}_{1}\cdot \vec{\beta}%
_{2} & \cdots & -\vec{\beta}_{1}\cdot \vec{\alpha}_{2N}+\vec{\alpha}%
_{1}\cdot \vec{\beta}_{2N} \\
\vdots & \vdots & \ddots & \vdots \\
-\vec{\beta}_{2N}\cdot \vec{\alpha}_{1}+\vec{\alpha}_{2N}\cdot \vec{\beta}%
_{1} & -\vec{\beta}_{2N}\cdot \vec{\alpha}_{2}+\vec{\alpha}_{2N}\cdot \vec{%
\beta}_{2} & \cdots & -\vec{\beta}_{2N}\cdot \vec{\alpha}_{2N}+\vec{\alpha}%
_{2N}\cdot \vec{\beta}_{2N}%
\end{array}%
\right)  \notag \\
&=&-2\pi \left(
\begin{array}{cccc}
l_{11} & l_{12} & \cdots & l_{1,2N} \\
\vdots & \vdots & \ddots & \vdots \\
l_{2N,1} & l_{2N,1} & \cdots & l_{2N,2N}%
\end{array}%
\right) .  \notag
\end{eqnarray}%
So we have%
\begin{equation*}
\det \left( W^{T}JW\right) =\left( \det W\right) ^{2}\det J=\left( -2\pi
\right) ^{2N}\det \left( l_{ss^{\prime }}\right) \equiv \left( -2\pi \right)
^{2N}\det L,
\end{equation*}%
\begin{equation*}
\left\vert \det J\right\vert =1\Longrightarrow \left\vert \det W\right\vert
=\left( 2\pi \right) ^{N}\sqrt{\left\vert \det L\right\vert },
\end{equation*}%
The volumes of three kinds of unit cells are given as follows
\begin{equation}
\upsilon _{3}=\frac{\left( 2\pi \right) ^{N}}{\sqrt{\left\vert \det
L\right\vert }},
\end{equation}%
\begin{equation*}
\upsilon _{2}=\left\vert \det \left(
\begin{array}{cc}
b & -\gamma \\
0 & a%
\end{array}%
\right) \right\vert =\left\vert \det b\det a\right\vert ,
\end{equation*}%
with $b_{ij}=\left( \vec{b}_{j}\right) _{i}.$ From (\ref{43}) we have
\begin{equation*}
\upsilon _{2}=\left( 2\pi \right) ^{N}.
\end{equation*}%
\begin{equation*}
\upsilon _{1}=\left\vert \det JW\right\vert =\left\vert \det W\right\vert
=\left( 2\pi \right) ^{N}\sqrt{\left\vert \det L\right\vert }.
\end{equation*}%
Thus we can obtain the ratios by volume among the three kinds of unit cells.%
\begin{equation*}
\frac{\upsilon _{1}}{\upsilon _{2}}=\sqrt{\left\vert \det L\right\vert },
\end{equation*}%
\begin{equation*}
\frac{\upsilon _{2}}{\upsilon _{3}}=\frac{\left( 2\pi \right) ^{N}}{\left(
2\pi \right) ^{N}\frac{1}{\sqrt{\left\vert \det L\right\vert }}}=\sqrt{%
\left\vert \det L\right\vert },
\end{equation*}%
\begin{equation*}
\sqrt{\left\vert \det L\right\vert }=\left\vert PffL\right\vert =N_{d}.
\end{equation*}%
This is the degree of degeneracy in $V_{1}$.

\section{Transformation of $\left\vert \vec{B},\vec{q}\right\rangle $ under
Group $G$ and Blocking Theorem}

\bigskip

In this section we discuss the transformation of the common eigenstates $%
\left\vert \vec{B},\vec{q}\right\rangle $ of $\left\{ U_{s}\right\} $ under
rotation $G$ . When $U_{s}=e^{iy_{s}},R_{k}\in G$ in consistent with (\ref%
{53}),
\begin{equation}
R_{k}^{-1}y_{s}R_{k}=y_{s}^{\prime }=\sum_{s^{\prime }}K_{ss^{\prime
}}y_{s^{\prime }}~~~K_{ss^{\prime }}\in
\mathbb{Z}
,~  \label{27}
\end{equation}%
\begin{equation}
\det K=1.
\end{equation}%
Thus, we have%
\begin{equation}
U_{s}^{\prime }=R_{k}^{-1}U_{s}R_{k}=\eta _{sk}\prod_{s^{\prime
}}U_{s^{\prime }}^{K_{ss^{\prime }}},
\end{equation}%
\begin{equation}
\eta _{sk}=\left( -1\right) ^{h_{sk}}=\pm 1.
\end{equation}%
From
\begin{equation*}
U_{s}\left\vert \vec{B},\vec{q}\right\rangle =\lambda _{s}\left\vert \vec{B},%
\vec{q}\right\rangle ,
\end{equation*}%
\begin{equation}
\lambda _{s}=\left( -1\right) ^{g_{s}}e^{i\left( \vec{\alpha}_{s}\cdot \vec{B%
}+\vec{\beta}_{s}\cdot \vec{q}\right) }\equiv \left( -1\right)
^{g_{s}}e^{i\mu _{s}},
\end{equation}%
we obtain%
\begin{eqnarray}
&&U_{s}R_{k}\left\vert \vec{B},\vec{q}\right\rangle =R_{k}\left(
R_{k}^{-1}U_{s}R_{k}\right) \left\vert \vec{B},\vec{q}\right\rangle  \notag
\\
&=&R_{k}\left( -1\right) ^{h_{sk}}\prod_{s^{\prime }}\lambda _{s^{\prime
}}^{K_{ss^{\prime }}}\left\vert \vec{B},\vec{q}\right\rangle  \notag \\
&=&\left( -1\right) ^{h_{sk}+\sum_{s^{\prime }}g_{s^{\prime }}K_{ss^{\prime
}}}e^{i\sum_{s^{\prime }}K_{ss^{\prime }}\left( \vec{\alpha}_{s^{\prime
}}\cdot \vec{B}+\vec{\beta}_{s^{\prime }}\cdot \vec{q}\right)
}R_{k}\left\vert \vec{B},\vec{q}\right\rangle .  \label{54}
\end{eqnarray}%
So $R_{k}\left\vert \vec{B},\vec{q}\right\rangle $ also belongs to the set
of eigenstates of $\left\{ U_{s}\right\} $, the eigenvalue of which is
\begin{eqnarray}
\lambda _{s}^{\prime } &=&\left( -1\right) ^{h_{sk}+\sum_{s^{\prime
}}g_{s^{\prime }}K_{ss^{\prime }}}e^{i\sum_{s^{\prime }}K_{ss^{\prime
}}\left( \vec{\alpha}_{s^{\prime }}\cdot \vec{B}+\vec{\beta}_{s^{\prime
}}\cdot \vec{q}\right) }  \label{25} \\
&\equiv &\left( -1\right) ^{g_{s}}e^{i\mu _{s}^{\prime }}.  \notag
\end{eqnarray}%
Define 2N-dimensional vectors $\vec{z},\vec{g},\vec{h}_{k}$ as
\begin{eqnarray*}
&&\{%
\begin{array}{c}
\left( \vec{z}\right) _{j}=B_{j} \\
\left( \vec{z}\right) _{N+j}=q_{j}%
\end{array}%
~,~j=1,2,\cdots ,N~\ ~ \\
&&\{%
\begin{array}{c}
\left( \vec{g}\right) _{s}=g_{s} \\
\left( \vec{h}_{k}\right) _{s}=h_{sk}%
\end{array}%
,~~s=1,2,\cdots ,2N
\end{eqnarray*}%
Recall the $2N\times 2N$ matrix $W$ is defined via $W_{s^{\prime }s}=\left(
\vec{w}_{s}\right) _{s^{\prime }},i.e.$%
\begin{eqnarray*}
&&\{%
\begin{array}{c}
W_{js}=\left( \vec{\alpha}_{s}\right) _{j} \\
W_{j+N,s}=\left( \vec{\beta}_{s}\right) _{j}%
\end{array}%
,~ \\
~j &=&1,2,\cdots .N~ \\
s &=&1,2,\cdots ,2N
\end{eqnarray*}%
Thus, we have
\begin{eqnarray}
K_{ss^{\prime }}\left( \vec{\alpha}_{s^{\prime }}\cdot \vec{B}+\vec{\beta}%
_{s^{\prime }}\cdot \vec{q}\right) &\equiv &K_{ss^{\prime }}\left( \alpha
_{js^{\prime }}B_{j}+~\beta _{js^{\prime }}q_{j}\right)  \notag  \label{24}
\\
&=&K_{ss^{\prime }}W_{s^{\prime \prime }s^{\prime }}z_{s^{\prime \prime }}
\notag \\
&=&\left( KW^{T}\vec{z}\right) _{s},
\end{eqnarray}%
In Eq.(\ref{25})%
\begin{eqnarray}
\mu _{s}^{\prime } &=&\left( KW^{T}\vec{z}\right) _{s}+\pi \left( \vec{h}%
_{k}+K\vec{g}-\vec{g}\right) _{s}  \notag \\
&=&\left[ W^{T}\left( \left( W^{T}\right) ^{-1}KW^{T}\vec{z}+\vec{\triangle}%
_{k}\right) \right] _{s}  \notag \\
&\equiv &\left[ W^{T}\left( K^{\prime }\vec{z}+\vec{\triangle}_{k}\right) %
\right] _{s},  \label{26}
\end{eqnarray}%
where $\vec{\triangle}_{k}=\pi \left( W^{T}\right) ^{-1}\left( \vec{h}%
_{k}+\left( K-1\right) \vec{g}\right) $ is a constant vector related to $%
R_{k}$. $\vec{\triangle}_{k}$ doesn't depend on $(\vec{B},\vec{q})$. We know
that the states are complete in the periodic unit cell $V_{2}$. The common
eigenstates of $\left\{ U_{s}\right\} $ with certain eigenvalue can be
spanned by $\left\{ \left\vert \left( \vec{B}_{0},\vec{q}_{0}\right) +\vec{t}%
_{l}\right\rangle \right\} _{l=1,2,\cdots ,N_{d}},$ where $\left( \vec{B}%
_{0},\vec{q}_{0}\right) $ is inside the nondegenerate little unit cell $%
\sigma _{00}$ which is made by $\left\{ \vec{\tau}_{s}\right\} $and $\vec{t}%
_{l}$ are degenerate lattice vectors in the periodic unit cell $V_{2}$.
\begin{equation}
\vec{t}_{l}=\sum_{s=1}^{2N}\vec{\tau}_{s}n_{ls},~~~\left( l=1,2,\cdots
,N_{d},~n_{ls}\in \mathbb{Z}\right)
\end{equation}%
For the common eigenstates $\left\vert \vec{B},\vec{q}\right\rangle $\ of $%
\left\{ U_{s}\right\} $ with same eigenvalue, their $\left( \vec{B},\vec{q}%
\right) $ at most differ by a degenerate lattice vector $\vec{t}_{l}$ So, we
obtain from (\ref{54})
\begin{equation}
R_{k}\left\vert \left( \vec{B}_{0},\vec{q}_{0}\right) +\vec{t}%
_{l}\right\rangle =\sum_{l^{\prime }=1}^{N_{d}}A_{l^{\prime }l}\left( \vec{B}%
_{0},\vec{q}_{0}\right) \left\vert \left( \vec{B}_{0}^{\prime },\vec{q}%
_{0}^{\prime }\right) +\vec{t}_{l^{\prime }}\right\rangle .  \label{21}
\end{equation}%
Here $\left( \vec{B}_{0},\vec{q}_{0}\right) $ and $\left( \vec{B}%
_{0}^{\prime },\vec{q}_{0}^{\prime }\right) $ both belong to the little unit
cell $\sigma _{00}$. Summation goes over the degenerate lattice sites in the
periodic unit cell $V_{2}$.

Let%
\begin{equation*}
\left( \vec{B}_{0},\vec{q}_{0}\right) =\vec{z}_{0},~~~\left( \vec{B}%
_{0}^{\prime },\vec{q}_{0}^{\prime }\right) =\vec{z}_{0}^{\prime }.
\end{equation*}%
Since $\lambda _{s}=\left( -1\right) ^{g_{s}}e^{i\left( W\vec{z}_{0}\right)
_{s}},~~~\lambda _{s}^{\prime }=\left( -1\right) ^{g_{s}}e^{i\left( W\vec{z}%
_{0}^{\prime }\right) _{s}},$ thus from (\ref{26}) we have
\begin{equation}
\vec{z}_{0}^{\prime }=K^{\prime }\vec{z}_{0}+\vec{\triangle}_{k}-\vec{t},
\label{22}
\end{equation}%
where $\vec{t}$ is a degenerate vector of $L_{3}$%
\begin{equation}
W^{T}\vec{t}=2\pi \vec{n},
\end{equation}%
\begin{equation*}
\left( \vec{n}\right) _{l}=n_{l}\in
\mathbb{Z}
,~~~l=1,2,\cdots ,2N.
\end{equation*}%
We call the mapping from $\vec{z}_{0}$ to $\vec{z}_{0}^{\prime }$ as mapping
induced by $R_{k}$ denoted by $\check{K}$ in the nondegenerate little unit
cell $\sigma _{00}$ . Consider an arbitrary degenerate vector $\vec{t}$, $%
\vec{t}^{\prime }=K^{\prime }\vec{t}=\left( W^{T}\right) ^{-1}KW^{T}\vec{t}%
=2\pi \left( W^{T}\right) ^{-1}K\vec{n}=2\pi \left( W^{T}\right) ^{-1}\vec{n}%
^{\prime }\Longrightarrow W^{T}\vec{t}^{\prime }=2\pi \vec{n}^{\prime }$,
namely $\vec{t}^{\prime }$ still belongs to the set of degenerate lattice
vectors. So mapping $K^{\prime }:\vec{z}^{\prime }=K^{\prime }\vec{z}$
change lattice $L_{3}$ into lattice $L_{3}.$ The original whole $\left( \vec{%
B},\vec{q}\right) $ space $X_{z}=\{\vec{z}=\vec{z}_{0}+\vec{t}\big\vert\vec{z%
}_{0}\in \sigma _{00},\vec{t}\in L_{3}\}$ can be built up by little unit
cell $\sigma _{00}$ according to degenerate lattice $L_{3}$. The linear
mapping $\check{K}^{\prime }$ on $X_{z}$ defined as $\check{K}^{\prime
}:z\rightarrow \vec{z}^{\prime }=K^{\prime }\vec{z}+\vec{\triangle}_{k}$
maps $X_{z}$ into $X_{z}$ and make $\sigma _{00}$ mapped into $\sigma
_{00}^{\prime }.$ Note that the original $X_{z}$ can be built up by $\sigma
_{00}$ according to degenerate lattice $L_{2},$ and
\begin{equation}
\det K^{\prime }=\det \left( \left( W^{T}\right) ^{-1}KW^{T}\right) =\det K=1
\label{23}
\end{equation}%
Since $\vec{t}^{\prime }=K^{\prime }\vec{t}\in L_{3}$ the mapping $K^{\prime
}$ maps degenerate lattice $L_{3}$ into $L_{3}$, so the space $\left( \vec{B}%
,\vec{q}\right) $ namely $X_{z}$ can also be built up by $\sigma
_{00}^{\prime }$ according to degenerate lattice $L_{3}$. Based on this fact
we infer that when $\sigma _{00}$ is generated by a finite number of $2N-1$
dimensional hyper planes (for example let $\sigma _{00}$ be hyper polyhedron
built by $\left\{ \vec{\tau}_{s}\right\} $), $\sigma _{00}$ and $\sigma
_{00}^{\prime }$ can be made up of the same little blocks $\sigma _{1}^{j}.$
This is because two kinds of periodic configurations respectively
corresponding to $\sigma _{00}$ and $\sigma _{00}^{\prime }$ are both able
to constitute $X_{z}$. The interfaces of $\sigma _{00}$ and $\sigma
_{00}^{\prime }$ in the configurations form and these little blocks
certainly can build up both $\sigma _{00}$ and $\sigma _{00}^{\prime }$. The
equations (\ref{21}) and (\ref{22}) imply that the mapping $\left( \vec{B}%
_{0},\vec{q}_{0}\right) \rightarrow \left( \vec{B}_{0}^{\prime },\vec{q}%
_{0}^{\prime }\right) $ can be realized by two steps: first map $\sigma
_{00} $ into $\check{K}^{\prime }\sigma _{00}=K^{\prime }\sigma _{00}+\vec{%
\triangle}_{k}=\sigma _{00}^{\prime }$, next according to some definite
block $\sigma _{1}^{j}$ drag it back to $\sigma _{00}$ by different
degenerate vectors $\vec{t}_{j}.$

Setting the preimage of $\sigma _{1}^{j}$ to be $\widetilde{\sigma }%
_{1}^{j}, $ we have
\begin{eqnarray*}
\sigma _{1}^{j} &=&K^{\prime }\widetilde{\sigma }_{1}^{j}+\vec{\triangle}%
_{k}+\vec{t}_{j} \\
&=&K^{\prime }\left( \widetilde{\sigma }_{1}^{j}+K^{\prime -1}\left( \vec{%
\triangle}_{k}+\vec{t}_{j}\right) \right) ,
\end{eqnarray*}%
where the $\sigma _{1}^{j}$ are smaller blocks inside $\sigma _{00},$ namely
$\sigma _{1}^{j}$ can be obtained from $\widetilde{\sigma }_{1}^{j}$ through
a linear mapping $\check{K}^{\prime (j)}:\vec{z}\rightarrow \vec{z}^{\prime
}=K^{\prime }\times \left( \vec{z}+\vec{c}_{j}\right) $, ($\vec{c}%
_{j}=K^{\prime -1}\left( \vec{\triangle}_{k}+\vec{t}_{j}\right) $). From (%
\ref{23}) we know this mapping is an volume-preserving mapping. So we obtain
a result: the effect of $R_{k}$ acting on the state vector $\left\vert
\left( \vec{B}_{0},\vec{q}_{0}\right) +\vec{t}_{l}\right\rangle $ is to
introduce a mapping $\left( \vec{B}_{0},\vec{q}_{0}\right) \rightarrow
\left( \vec{B}_{0}^{\prime },\vec{q}_{0}^{\prime }\right) $ as (\ref{21}).
This mapping can be described as follows: First cut $\sigma _{00}$ into
finite blocks $\widetilde{\sigma }_{1}^{j}$, which are then respectively
linearly mapped into $\sigma _{1}^{j},$ the shift vector $\vec{c}_{j}$ may
depend on $j$ and be different.

\bigskip

Next, through the following \textbf{Lemma 1} and \textbf{Lemma 2} we will
prove a \textbf{Blocking Theorem} about the mapping $\left(
B_{0},q_{0}\right) \xrightarrow[\check{K}]{R\in G}\left( B_{0}^{\prime
},q_{0}^{\prime }\right) $

\begin{lemma}
For the mapping $\check{K}^{\prime j}\vec{z}=K^{\prime }\times \left( \vec{z}%
+\vec{c}^{j}\right) ,K^{\prime }\neq I$ we can always divide $X_{z}$ namely
the space of $\left( \vec{B},\vec{q}\right) $ into a finite number of
sectorial regions $V^{j}$. Under the mapping different sectors exchange with
each other and the image of each block doesn't overlap itself. The meaning
of "overlap" is that there exist a common $2N$-dimensional little continuous
regions $V^{j}\cap V^{j^{\prime }}.$
\end{lemma}

\begin{proof}
First let the fixed point of the mapping be at the origin of coordinates.

Matrix $K^{\prime }$($2N\times 2N$) has the following property:

\begin{equation*}
\left( K^{\prime }\right) ^{n_{k}}=I.
\end{equation*}%
($G$ is a finite group, $R_{k}^{n_{k}}=id$)The rank of $G$ is finite and $%
K^{\prime }$ is a $2N$-dimensional representation of a cyclic group. Based
on the theorem about finite group that if the representation is reducible,
surely completely reducible, and a cyclic group only has one-dimensional
irreducible representation, after reduction Matrix $K^{\prime }$ goes to
\begin{equation*}
\left(
\begin{array}{cccc}
\lambda _{1} &  &  &  \\
& \lambda _{2} &  &  \\
&  & \ddots &  \\
&  &  & \lambda _{2N}%
\end{array}%
\right)
\end{equation*}%
The corresponding invariant sub-space is one-dimensional, namely for
eigenvector $\psi _{j},K^{\prime }\psi _{j}=\lambda _{j}\psi _{j},\lambda
_{j}^{n_{k}}=1$, so $\lambda _{j}$ is a $n_{k}$ order root of unity. If $%
\lambda _{j}$ equals to $1$, $\psi _{j}$ can be chosen as a real vector,
which is an invariant direction under action of $K^{\prime }$. If $\lambda
_{j}$ is an imaginary number $e^{i\omega _{j}}$, $\psi _{j}$ can be
decomposed into $\phi _{1}+i\phi _{2}$, here $\phi _{1}$ and $\phi _{2}$ are
real vectors.
\begin{eqnarray}
K^{\prime }\left( \phi _{1}+i\phi _{2}\right) &=&\left( \cos \omega +i\sin
\omega \right) \left( \phi _{1}+i\phi _{2}\right)  \notag \\
&=&\left( \phi _{1}\cos \omega -\phi _{2}\sin \omega \right) +i\left( \phi
_{2}\cos \omega +\phi _{1}\sin \omega \right) .  \label{51}
\end{eqnarray}%
Since $K^{\prime }$ is real, we have%
\begin{eqnarray*}
&\Longrightarrow &K^{\prime }\phi _{1}=\phi _{1}\cos \omega -\phi _{2}\sin
\omega , \\
&&K^{\prime }\phi _{2}=\phi _{2}\cos \omega +\phi _{1}\sin \omega .
\end{eqnarray*}%
This is a rotation in the plane spanned by $\phi _{1}$ and $\phi _{2}.$ The
complex conjugate of (\ref{51}) implies that the eigenvalue $\lambda
=e^{-i\omega _{j}}$ corresponding to $\psi ^{\ast }$ corresponds to the same
plane too, which is an invariant plane because $K^{\prime }$ always maps $%
\phi _{1}$ and $\phi _{2}$ to their linear combination. Since $K^{\prime }$
is completely reducible, we can prove that we always have a co-subspace
which is also invariant.

Let the invariant plane be $X^{\prime }$ and the co-subspace be
$X^{\prime \prime }.$ Any vector $\vec{z}\in X_{z}$ can be
uniquely decomposed as
\begin{equation*}
\vec{z}=\vec{z}_{1}+\vec{z}_{2},~~\vec{z}_{1}\in X^{\prime },\vec{z}_{2}\in
X^{\prime \prime }
\end{equation*}%
The above discussion means%
\begin{equation*}
K^{\prime }\vec{z}=K^{\prime }(\vec{z}_{1}+\vec{z}_{2})=K^{\prime }\vec{z}%
_{1}+K^{\prime }\vec{z}_{2},
\end{equation*}%
where $K^{\prime }\vec{z}_{1}\in X^{\prime },K^{\prime }\vec{z}_{2}\in
X^{\prime \prime }.$Let $\vec{c}^{j}=\vec{c}_{1}^{j}+\vec{c}_{2}^{j},\vec{c}%
_{1}^{j}\in X^{\prime },\vec{c}_{2}^{j}\in X^{\prime \prime },$we have%
\begin{equation*}
\check{K}^{\prime }\vec{z}=K^{\prime }(\vec{z}_{1}+\vec{c}%
_{1}^{j})+K^{\prime }(\vec{z}_{2}+\vec{c}_{2}^{j})
\end{equation*}%
with
\begin{equation*}
K^{\prime }(\vec{z}_{1}+\vec{c}_{1}^{j})\in X^{\prime },K^{\prime }(\vec{z}%
_{2}+\vec{c}_{2}^{j})\in X^{\prime \prime }.
\end{equation*}%
Thus the mapping $\check{K}^{\prime }$ induces a map in $X^{\prime
},$ which is a shift $\vec{c}_{1}^{j}$ that follows a rotation by
equation (\ref{51}). We can always divide $X^{\prime }$ into
$n_{k}$ sector by lines in $X^{\prime }, $ such that after mapping
each sector doesn't overlap itself. These sectors together with
$X^{\prime \prime }$ form division in $X_{z}$, which has the
desired property. Concretely, let
\begin{equation*}
X^{\prime }=D_{1}\cup D_{2}\cdots \cup D_{n_{k}},\vec{z}=\vec{z}_{1}+\vec{z}%
_{2},\vec{z}_{1}\in D_{l}
\end{equation*}%
\begin{equation*}
X_{z}=D_{1}\oplus X^{\prime \prime }\cup D_{2}\oplus X^{\prime \prime
}\cdots \cup D_{n_{k}}\oplus X^{\prime \prime }
\end{equation*}%
\begin{equation*}
K^{\prime }(\vec{z}_{1}+\vec{c}_{1}^{j})\in D_{m},m\neq l
\end{equation*}%
\begin{equation*}
\Longrightarrow \check{K}^{\prime }\vec{z}=K^{\prime }(\vec{z}_{1}+\vec{c}%
_{1}^{j})+K^{\prime }(\vec{z}_{2}+\vec{c}_{2}^{j})=\vec{r}_{1}+\vec{r}_{2}.
\end{equation*}%
\begin{equation*}
\vec{r}_{1}\in D_{m},\vec{r}_{2}\in X^{\prime \prime }\Longrightarrow \vec{r}%
_{1}+\vec{r}_{2}\in D_{m}\oplus X^{\prime \prime },
\end{equation*}%
which doesn't overlap $D_{l}\oplus X^{\prime \prime }.$The lemma
is valid.
\end{proof}

\begin{lemma}
For group $G$ (or its any subgroup $G^{j}$), to consider any figure point
set of $\sigma _{00},$ we can always get a new figure which contains the
original figure (point set), and the figure is invariant under action of $G.$
\end{lemma}

\begin{proof}
Let any point $P_{1}$ be mapped into $P_{1},P_{2},\cdots ,P_{\left\vert
G\right\vert },$ then the point set $\left\{ P_{1},P_{2},\cdots
,P_{\left\vert G\right\vert }\right\} $ doesn't change. So an arbitrary
point set $S_{1}$ is mapped by $R_{k}\in G$ into $S_{1},S_{2},\cdots
,S_{\left\vert G\right\vert },~$then $\cup _{j}S_{j}$ is an invariant point
set.
\end{proof}

\begin{theorem}[Blocking Theorem]
We can always divide the little unit cell$\ \sigma _{00}$ into a finite
number of blocks $\sigma _{3}^{q}$ such that they exchange positions with
each other under transformation $R_{k}\in G$: $\left( \vec{B}_{0},\vec{q}%
_{0}\right) \longrightarrow \left( \vec{B}_{0}^{\prime },\vec{q}_{0}^{\prime
}\right) $ and make their image not overlap themselves for any $R_{k}\neq
id. $
\end{theorem}

\begin{proof}
Consider the mapping induced by $R_{k},$ which maps the subset of $\sigma
_{00}:\widetilde{\sigma }_{1}^{j}\rightarrow \sigma _{1}^{j},$ $\left\{
\widetilde{\sigma }_{1}^{j}\right\} $ and $\left\{ \sigma _{1}^{j}\right\} $
can both build up $\sigma _{00}.$ Meanwhile the mapping $\check{K}^{\prime
\left( j\right) }$ :$\widetilde{\sigma }_{1}^{j}\rightarrow \sigma _{1}^{j}$
is a linear mapping, which is in general not homogeneous. Thanks to $\mathbf{%
Lemma~1}$, we can divide $\widetilde{\sigma }_{1}^{j}$ into several little
sections $\widetilde{\sigma }_{2}^{p},$ the image of each little section
doesn't overlap itself under $\check{K}^{\prime \left( j\right) }$ . This
kind of division together with their image (namely $\sigma _{2}^{p}$ the
division of $\sigma _{1}^{j}$,) form a figure by the boundaries of the
regions. We do this procedure for every $R\in G,$ and finally get a figure.
Based on $\mathbf{Lemma~2}$, \ we add up the divisions induced by every
group element to obtain the final figure which is invariant under action of
any group element $R_{k}\in G$. At last we obtain a division of $\sigma
_{00},$ named as $\sigma _{3}^{q},$ which only exchange their position under
the action of $G,$ since the whole figure keep invariant under $G$ and the
image of any little block under $R_{k}\in G$ doesn't overlap itself, which
is because each block included in the figure certainly belongs to one block
of the division $\left\{ \widetilde{\sigma }_{2}^{p}\right\} $ corresponding
to certain definite $R_{k},$ thus $R_{k}$\ must shift its position due to $%
\mathbf{Lemma~1}.$
\end{proof}

\bigskip

Based on the \textbf{Blocking Theorem}, we have the following corollary:

\begin{corollary}
If the mapping $\sigma _{3}^{q}\rightarrow \sigma _{3.}^{q\prime }$ given by
$R_{k_{1}}$ and $R_{k_{2}}$ are the same, then $R_{k_{1}}=R_{k_{2}}$.
\end{corollary}

\begin{proof}
If not, the mapping given by $R_{k_{1}}$ and $R_{k_{2}}^{-1}$ will make $%
\sigma _{3}^{q}$ overlap itself, however it is possible only when $%
R_{k_{1}}\cdot R_{k_{2}}^{-1}=1$. The conclusion follows.
\end{proof}

\bigskip

\begin{corollary}
We can always divide $\sigma _{00}$ into a collection of $\left\vert
G\right\vert $ little blocks $\sigma _{3}^{q}$($\left\vert G\right\vert $ is
the~rank~of~group~$G$ ), i.e. $\left\{ S^{l}\right\} (l=1,2,\cdots
,\left\vert G\right\vert ),\bigcup\limits_{l}S^{l}=$ $\sigma _{00}$, under $%
R_{k}\in G,$ $S^{1}$ can be mapped into any set $S^{k},k=1,2,\cdots
,\left\vert G\right\vert .$
\end{corollary}

\begin{proof}
Arbitrarily to take a little block $\sigma _{3}^{q}$, each transformation
induced by the group element of $G$ can change it into $\sigma
_{3}^{q}=\sigma _{3}^{q_{1}},\sigma _{3}^{q_{2}},\cdots ,\sigma
_{3}^{q_{\left\vert G\right\vert }}$, There is no superposition for these
blocks due to \ corollary 1. Using the above method, pick out $\left\vert
G\right\vert $ $\sigma _{3}^{q}$ and deal with the rest similarly. Since the
number of the blocks is finite, the selection can always be finished.
Therefore the $S^{1}$ can be made up of the first blocks picking out every
time, the corollary is obtained.
\end{proof}

\bigskip

\begin{corollary}
Let $\sigma _{4}\subset \sigma _{3}^{q_{j}}$ be a continuous region
belonging to $X_{z},$ its dimension is $2N,$ if the mapping by both $%
R_{k_{1}}$ and $R_{k_{2}}$ send $\sigma _{4}$ to the same continuous region $%
\sigma _{4}^{\prime }\subset \sigma _{3}^{q_{j}^{\prime }},$ then we have
\begin{equation}
R_{k_{1}}\left\vert \left( \vec{B}_{0},\vec{q}_{0}\right) +\vec{t}%
_{l}\right\rangle =R_{k_{2}}\left\vert \left( \vec{B}_{0},\vec{q}_{0}\right)
+\vec{t}_{l}\right\rangle .  \label{34}
\end{equation}
\end{corollary}

\begin{proof}
Due to Lemma 1, $R_{k_{1}}=R_{k_{2}}.$
\end{proof}

\bigskip

\section{ The Complete Set of the Projectors of T$^{2N}/G$}

Next we will construct the complete set of the projectors on $T^{2N}/G$. Let
$\hat{O}$ be an operator on $T^{2N}$ and $\hat{O}$ commutates with $\left\{
U_{s}\right\} .$ Let $\left\vert \psi \right\rangle $ be the common
eigenstate of $\left\{ U_{s}\right\} $, then $\hat{O}\left\vert \psi
\right\rangle $ is also their common eigenstate.%
\begin{equation*}
U_{s}\hat{O}\left\vert \psi \right\rangle =\hat{O}U_{s}\left\vert \psi
\right\rangle =\lambda _{s}\hat{O}\left\vert \psi \right\rangle ,
\end{equation*}

namely $\hat{O}\left\vert \psi \right\rangle $ is also the eigenstate of $%
U_{s}$ with the same eigenvalue as for $\left\vert \psi \right\rangle $. So
we have%
\begin{equation}
\hat{O}\left\vert \left( \vec{B}_{0},\vec{q}_{0}\right) +\vec{t}%
_{l}\right\rangle =\sum_{l^{\prime }}M_{l^{\prime }l}^{0}\left\vert \left(
\vec{B}_{0},\vec{q}_{0}\right) +\vec{t}_{l^{\prime }}\right\rangle ,
\label{52}
\end{equation}%
$M_{l^{\prime }l}^{0}$ is a function with respect to $\left( \vec{B}_{0},%
\vec{q}_{0}\right) .$Obviously,%
\begin{eqnarray}
\hat{A}\hat{B}\left\vert \left( \vec{B}_{0},\vec{q}_{0}\right) +\vec{t}%
_{l^{\prime }}\right\rangle &=&\hat{A}\sum_{l^{\prime }}M_{l^{\prime
}l}^{B}\left\vert \left( \vec{B}_{0},\vec{q}_{0}\right) +\vec{t}_{l^{\prime
}}\right\rangle  \notag \\
&=&\sum_{l^{\prime }l^{\prime \prime }}M_{l^{\prime }l}^{B}M_{l^{\prime
\prime }l^{\prime }}^{A}\left\vert \left( \vec{B}_{0},\vec{q}_{0}\right) +%
\vec{t}_{l^{\prime \prime }}\right\rangle  \notag \\
&=&\left( M^{A}M^{B}\right) _{l^{\prime \prime }l}\left\vert \left( \vec{B}%
_{0},\vec{q}_{0}\right) +\vec{t}_{l^{\prime \prime }}\right\rangle .
\end{eqnarray}%
Let $M^{P}\left( \vec{B}_{0}~\vec{q}_{0}\right) $\ be the corresponding $%
N_{d}\times N_{d}$ matrix, then for every projector of $T^{2N},$ we have%
\begin{equation*}
P\rightarrow M^{P}\left( \vec{B}_{0}~\vec{q}_{0}\right) ,\text{ }
\end{equation*}%
\begin{equation*}
\left( M^{P}\right) ^{2}=M^{P}.
\end{equation*}%
Since $\left\vert \left( \vec{B}_{0},\vec{q}_{0}\right) +\vec{t}_{l^{\prime
\prime }}\right\rangle $ is orthogonal and complete, we may rewrite (\ref{15}%
) as
\begin{equation*}
id=\frac{1}{\upsilon }\sum\limits_{l=1}^{N_{d}}\int_{\sigma _{_{00}}}d^{N}%
\vec{B}_{0}d^{N}\vec{q}_{0}\left\vert \left( \vec{B}_{0},\vec{q}_{0}\right) +%
\vec{t}_{l}\right\rangle \left\langle \left( \vec{B}_{0},\vec{q}_{0}\right) +%
\vec{t}_{l}\right\vert .
\end{equation*}%
Thus any operator $\hat{O}$ can be expressed by their corresponding $%
M_{l^{\prime }l}^{0}\left( \vec{B}_{0},\vec{q}_{0}\right) $ as
\begin{eqnarray}
\hat{O} &=&\frac{1}{v}\sum_{l}\int_{\sigma _{00}}d^{N}B_{0}d^{N}q_{0}\hat{O}%
\left\vert \left( \vec{B}_{0},\vec{q}_{0}\right) +\vec{t}_{l}\right\rangle
\left\langle \left( \vec{B}_{0},\vec{q}_{0}\right) +\vec{t}_{l}\right\vert
\notag \\
&=&\frac{1}{v}\sum_{ll^{\prime }}\int_{\sigma
_{00}}d^{N}B_{0}d^{N}q_{0}M_{l^{\prime }l}^{0}\left( \vec{B}_{0},\vec{q}%
_{0}\right) \left\vert \left( \vec{B}_{0},\vec{q}_{0}\right) +\vec{t}%
_{l^{\prime }}\right\rangle \left\langle \left( \vec{B}_{0},\vec{q}%
_{0}\right) +\vec{t}_{l}\right\vert .  \notag\\ \label{50}
\end{eqnarray}

Conversely, for a given $M_{l^{\prime }l}^{0}\left( \vec{B}_{0},\vec{q}%
_{0}\right) $ we can construct an operator $\hat{O}$ via the right-hand side
of (\ref{50}). One can show that the operator satisfies $[\hat{O}%
,U_{s}]=0,(s=1,2,\cdots ,2N)$\ and the equation (\ref{52}).

Next we study the relation between the $M\left( \vec{B}_{0},\vec{q}%
_{0}\right) $ matrices before and after action of rotation $R.$ Recall (\ref%
{21}), we have
\begin{eqnarray}
\left\vert \psi \right\rangle &=&R^{-1}\hat{O}R\left\vert \left( \vec{B}_{0},%
\vec{q}_{0}\right) +\vec{t}_{l}\right\rangle =R^{-1}\hat{O}A_{l^{\prime
}l}\left( \vec{B}_{0},\vec{q}_{0}\right) \left\vert \left( \vec{B}%
_{0}^{\prime },\vec{q}_{0}^{\prime }\right) +\vec{t}_{l^{\prime
}}\right\rangle  \notag \\
&=&\sum_{l^{\prime }l^{\prime \prime }}A_{l^{\prime }l}\left( \vec{B}_{0},%
\vec{q}_{0}\right) M_{l^{\prime \prime }l^{\prime }}\left( \vec{B}%
_{0}^{\prime },\vec{q}_{0}^{\prime }\right) R^{-1}\left\vert \left( \vec{B}%
_{0}^{\prime },\vec{q}_{0}^{\prime }\right) +\vec{t}_{l^{\prime \prime
}}\right\rangle .  \label{49}
\end{eqnarray}%
Let
\begin{equation*}
R^{-1}\left\vert \left( \vec{B}_{0}^{\prime },\vec{q}_{0}^{\prime }\right) +%
\vec{t}_{l^{\prime \prime }}\right\rangle =\sum_{k}A_{R^{-1}}\left( \vec{B}%
_{0}^{\prime },\vec{q}_{0}^{\prime }\right) _{kl^{\prime \prime }}\left\vert
\left( \vec{B}_{0}^{\prime \prime },\vec{q}_{0}^{\prime \prime }\right) +%
\vec{t}_{k}\right\rangle
\end{equation*}%
Because when $\hat{O}=id,$we have
\begin{eqnarray*}
R^{-1}IR\left\vert \left( \vec{B}_{0},\vec{q}_{0}\right) +\vec{t}%
_{l}\right\rangle &=&\sum_{kl^{\prime }l^{\prime \prime }}A_{l^{\prime
}l}\left( \vec{B}_{0},\vec{q}_{0}\right) \delta _{l^{\prime \prime
}l^{\prime }}\left( A_{R^{-1}}\left( \vec{B}_{0}^{\prime },\vec{q}%
_{0}^{\prime }\right) \right) _{kl^{\prime \prime }}\left\vert \left( \vec{B}%
_{0}^{\prime \prime },\vec{q}_{0}^{\prime \prime }\right) +\vec{t}%
_{k}\right\rangle \\
&=&\left\vert \left( \vec{B}_{0},\vec{q}_{0}\right) +\vec{t}_{l}\right\rangle
\\
&\Longrightarrow &\left( \vec{B}_{0}^{\prime \prime },\vec{q}_{0}^{\prime
\prime }\right) =\left( \vec{B}_{0},\vec{q}_{0}\right) , \\
A_{R^{-1}}\left( \vec{B}_{0}^{\prime },\vec{q}_{0}^{\prime }\right) &=&\left[
A\left( \vec{B}_{0},\vec{q}_{0}\right) \right] ^{-1}\equiv A^{-1}\left( \vec{%
B}_{0},\vec{q}_{0}\right) .
\end{eqnarray*}

So, Eq.(\ref{49}) becomes%
\begin{equation}
\left\vert \psi \right\rangle =\left[ A^{-1}\left( \vec{B}_{0},\vec{q}%
_{0}\right) M\left( \vec{B}_{0}^{\prime },\vec{q}_{0}^{\prime }\right)
A\left( \vec{B}_{0},\vec{q}_{0}\right) \right] _{kl}\left\vert \left( \vec{B}%
_{0},\vec{q}_{0}\right) +\vec{t}_{k}\right\rangle ,
\end{equation}%
namely, when $\hat{O}$ corresponds to $M,$ $R^{-1}\hat{O}R$ corresponds to
\begin{equation*}
\widetilde{M}\left( \vec{B}_{0},\vec{q}_{0}\right) =A^{-1}\left( \vec{B}_{0},%
\vec{q}_{0}\right) M\left( \vec{B}_{0}^{\prime },\vec{q}_{0}^{\prime
}\right) A\left( \vec{B}_{0},\vec{q}_{0}\right) .
\end{equation*}

We have
\begin{equation}
M\left( \vec{B}_{0}^{\prime },\vec{q}_{0}^{\prime }\right) =A\left( \vec{B}%
_{0},\vec{q}_{0}\right) \widetilde{M}\left( \vec{B}_{0},\vec{q}_{0}\right)
A^{-1}\left( \vec{B}_{0},\vec{q}_{0}\right) .
\end{equation}

Then in order to make $\hat{O}$ invariant under $R,$ we need the sufficient
and necessary condition
\begin{equation}
\widetilde{M}=M  \label{48}
\end{equation}%
as a matrix function of $\left( \vec{B}_{0},\vec{q}_{0}\right) .$\ Namely,
\begin{equation}
A\left( \vec{B}_{0},\vec{q}_{0}\right) M\left( \vec{B}_{0},\vec{q}%
_{0}\right) A^{-1}\left( \vec{B}_{0},\vec{q}_{0}\right) =M\left( \vec{B}%
_{0}^{\prime },\vec{q}_{0}^{\prime }\right) .  \label{55}
\end{equation}

In the following, we begin to construct the complete set of the operators $%
P: $

First of all, in a little region $S^{1}$ of $\sigma _{00},$ one matrix $M(%
\vec{B}_{0},\vec{q}_{0})$ satisfying $M^{2}=M$ can certainly be generated by
the general formula $M(\vec{B}_{0},\vec{q}_{0})=T^{-1}(\vec{B}_{0},\vec{q}%
_{0})M_{0}(\vec{B}_{0},\vec{q}_{0})T(\vec{B}_{0},\vec{q}_{0})$(ref.\cite%
{Deng}), where $M_{0}(\vec{B}_{0},\vec{q}_{0})$ is an arbitrary diagonal
matrix with diagonal elements taking $0$ or $1,$ and $T$ is an arbitrary
invertible matrix. $T$ and $M_{0}$ are both functions with respect to $(\vec{%
B}_{0},\vec{q}_{0}).$ Through the equation (\ref{55}) we assign $M$ in the
region $S^{k}$ as%
\begin{equation}
M\left( \vec{B}_{0}^{\prime },\vec{q}_{0}^{\prime }\right) =A_{R_{k}}\left(
\vec{B}_{0},\vec{q}_{0}\right) M\left( \vec{B}_{0},\vec{q}_{0}\right)
A_{R_{k}}^{-1}\left( \vec{B}_{0},\vec{q}_{0}\right) ,
\end{equation}%
where $\left( \vec{B}_{0}^{\prime },\vec{q}_{0}^{\prime }\right) $ is the
image of $\left( \vec{B}_{0},\vec{q}_{0}\right) $ induced by $R_{k}\in G,$ $%
A $ is the matrix corresponding to $R_{k}.$ By corollary 2, this
construction can cover the whole cell $\sigma _{00}.$ We thus get $M$
corresponding to the whole $\sigma _{00}.$ The $M\left( \vec{B}_{0},\vec{q}%
_{0}\right) $ in the whole $\sigma _{00}$ keeps invariant under action of
any group element $R\in G$. The following is the proof:

\begin{proof}
Let%
\begin{equation}
\widetilde{M}\left( \vec{B}_{0}^{\prime },\vec{q}_{0}^{\prime }\right)
=A\left( \vec{B}_{0},\vec{q}_{0}\right) M\left( \vec{B}_{0},\vec{q}%
_{0}\right) A^{-1}\left( \vec{B}_{0},\vec{q}_{0}\right) .  \label{44}
\end{equation}%
where $R$ maps $\left( \vec{B}_{0},\vec{q}_{0}\right) $ to $\left( \vec{B}%
_{0}^{\prime },\vec{q}_{0}^{\prime }\right) .$

\begin{enumerate}
\item If $\left( \vec{B}_{0},\vec{q}_{0}\right) \in S^{1},$ according to the
construction, the conclusion is obviously valid.

\item If $\left( \vec{B}_{0},\vec{q}_{0}\right) \notin S^{1}$, set $\left(
\vec{B}_{0},\vec{q}_{0}\right) \in S^{^{\prime }},\left( \vec{B}_{0},\vec{q}%
_{0}\right) =R^{\prime }\left( \tilde{\vec{B}},\tilde{\vec{q}}\right) $
where $R^{\prime }\left( \tilde{\vec{B}},\tilde{\vec{q}}\right) $ is the
mapping from \ point $\left( \tilde{\vec{B}},\tilde{\vec{q}}\right) $ in the
$S^{1}$ to $\left( \vec{B}_{0},\vec{q}_{0}\right) ,$ then under $R$, from (%
\ref{44}) we have%
\begin{eqnarray}
\widetilde{M}\left( \vec{B}_{0}^{\prime },\vec{q}_{0}^{\prime }\right)
&=&A_{R}\left( \vec{B}_{0},\vec{q}_{0}\right) M\left( \vec{B}_{0},\vec{q}%
_{0}\right) A_{R}^{-1}\left( \vec{B}_{0},\vec{q}_{0}\right)  \notag
\label{45} \\
&=&A_{R}\left( \vec{B}_{0},\vec{q}_{0}\right) A_{R^{\prime }}\left( \tilde{%
\vec{B}},\tilde{\vec{q}}\right) M\left( \tilde{\vec{B}},\tilde{\vec{q}}%
\right) A_{R^{\prime }}^{-1}\left( \tilde{\vec{B}},\tilde{\vec{q}}\right)
A_{R}^{-1}\left( \vec{B}_{0},\vec{q}_{0}\right) .  \notag \\
&&
\end{eqnarray}

Since under action of the group elements of $G$, the transformation of $%
\left\vert \left( \vec{B}_{0},\vec{q}_{0}\right) +\vec{t}_{l}\right\rangle $
should form a representation of $G,$ we have (\ref{21})
\begin{equation*}
A_{R}\left( \vec{B}_{0},\vec{q}_{0}\right) A_{R^{\prime }}\left( \tilde{\vec{%
B}},\tilde{\vec{q}}\right) =A_{(RR^{\prime })}\left( \tilde{\vec{B}},\tilde{%
\vec{q}}\right) .
\end{equation*}%
The mapping induced by $\left( RR^{\prime }\right) $ in the $\sigma _{00}$
just send $\left( \tilde{\vec{B}},\tilde{\vec{q}}\right) \rightarrow \left(
\vec{B}_{0}^{\prime },\vec{q}_{0}^{\prime }\right) :$%
\begin{eqnarray}
RR^{\prime }\left\vert \left( \tilde{\vec{B}},\tilde{\vec{q}}\right) +\vec{t}%
_{l_{1}}\right\rangle &=&RA_{R^{\prime }}\left( \tilde{\vec{B}},\tilde{\vec{q%
}}\right) _{l_{2}l_{1}}\left\vert \left( \vec{B}_{0},\vec{q}_{0}\right) +%
\vec{t}_{l_{2}}\right\rangle  \notag \\
&=&A_{R^{\prime }}\left( \tilde{\vec{B}},\tilde{\vec{q}}\right)
_{l_{2}l_{1}}A_{R}\left( \vec{B}_{0},\vec{q}_{0}\right)
_{l_{3}l_{2}}\left\vert \left( \vec{B}_{0}^{\prime },\vec{q}_{0}^{\prime
}\right) +\vec{t}_{l_{3}}\right\rangle  \notag \\
&=&A_{\left( RR^{\prime }\right) }\left( \tilde{\vec{B}},\tilde{\vec{q}}%
\right) _{l_{3}l_{1}}\left\vert \left( \vec{B}_{0}^{\prime },\vec{q}%
_{0}^{\prime }\right) +\vec{t}_{l_{3}}\right\rangle .  \label{46}
\end{eqnarray}%
So, we get%
\begin{equation}
A_{R}\left( \vec{B}_{0},\vec{q}_{0}\right) A_{R^{\prime }}\left( \tilde{\vec{%
B}},\tilde{\vec{q}}\right) =A_{\left( RR^{\prime }\right) }\left( \tilde{%
\vec{B}},\tilde{\vec{q}}\right)
\end{equation}%
From (\ref{45}) we have%
\begin{equation}
\widetilde{M}\left( \vec{B}_{0}^{\prime },\vec{q}_{0}^{\prime }\right)
=A_{\left( RR^{\prime }\right) }\left( \tilde{\vec{B}},\tilde{\vec{q}}%
\right) M\left( \tilde{\vec{B}},\tilde{\vec{q}}\right) A_{\left( RR^{\prime
}\right) }^{-1}\left( \tilde{\vec{B}},\tilde{\vec{q}}\right) .  \label{47}
\end{equation}
\end{enumerate}
\end{proof}

From (\ref{46}) we see that $RR^{\prime }$ maps $S^{1\text{ }}$to $S^{\prime
}$ which implies $\widetilde{M}\left( \vec{B}_{0}^{\prime },\vec{q}%
_{0}^{\prime }\right) =M\left( \vec{B}_{0}^{\prime },\vec{q}_{0}^{\prime
}\right) .$ Thus based on the above construction, the right hand side of (%
\ref{47}) equals to $M\left( \vec{B}_{0}^{\prime },\vec{q}_{0}^{\prime
}\right) $. Since $M$ constructed like this keeps invariant under action of
any $R\in G$, the corresponding operator formed as (\ref{50}) is also
invariant under $G$. Due to $M^{2}=M$ in the region $S^{1},$ we get $%
\widetilde{M}^{2}=\widetilde{M}$. By construction, therefore $M^{2}=M$ also
holds in the other regions $S^{k}$.

So, we achieve the complete set of projectors on the $T^{2N}$\bigskip $/G$.

\section{\protect\bigskip Discussion}

When we take
\begin{equation}
M\left( \vec{B}_{0},\vec{q}_{0}\right) _{l_{2}l_{1}}=\frac{<\left( \vec{B}%
_{0},\vec{q}_{0}\right) +\vec{t}_{l_{2}}|\Omega ><\Omega |\left( \vec{B}_{0},%
\vec{q}_{0}\right) +\vec{t}_{l_{1}}>}{\sum_{l=1}^{N_{d}}<\left( \vec{B}_{0},%
\vec{q}_{0}\right) +\vec{t}_{l}|\Omega ><\Omega |\left( \vec{B}_{0},\vec{q}%
_{0}\right) +\vec{t}_{l}>}
\end{equation}%
and when $R_{k}\left\vert \Omega \right\rangle =e^{i\omega _{k}}\left\vert
\Omega \right\rangle ,$ $M\left( \vec{B}_{0},\vec{q}_{0}\right) $ satisfies
all the requirements about the reduced matrix $M(\vec{B}_{0},\vec{q}_{0})$
of the projector on the $T^{2N}$\bigskip $/G,$ This is the $GHS$
construction \cite{GHS}. We can construct the closed solution of the
projectors in terms of hyper elliptic function, we will study this question
in the sequel.

\section{Acknowledgments}
    This work is supported by the National Natural Science Foundation of China granted by
    No.10175050.

\end{document}